\documentclass[sigconf,usenames,dvipsnames]{acmart}
\usepackage{courier}
\usepackage{url}
\usepackage{balance}
\usepackage{enumitem}
\usepackage{caption}	
\usepackage{multirow}
\usepackage{multicol}
\usepackage{amsmath}
\usepackage{float}
\usepackage[normalem]{ulem}
\usepackage{xspace}
\usepackage{listings}
\usepackage{subcaption}
\usepackage{hyperref}

\newcommand{\eg}{\textit{e.g.,}\xspace}

\newcommand{\etal}{\textit{et al.}\xspace}

\newcommand{\fusion}{{\sc Fusion}\xspace}
\newcommand{\crashscope}{{\sc CrashScope}\xspace}
\newcommand{\MonkeyLab}{{\sc MonkeyLab}\xspace}

\lstdefinestyle{interfaces}{
  float,
  floatplacement=tb
}

\include{macro}

\begin{document}

\title{Overcoming Language Dichotomies: Toward Effective Program\\ Comprehension for Mobile App Development}
\author{Kevin Moran}
\affiliation{\institution{College of William and Mary} \department{Department of Computer Science}}
\email{kpmoran@cs.wm.edu}
\author{Carlos Bernal-C\'{a}rdenas}
\affiliation{\institution{College of William and Mary} \department{Department of Computer Science}}
\email{cebernal@cs.wm.edu}
\author{Mario Linares-V\'{a}squez}
\affiliation{\institution{Universidad de los Andes} \department{Systems and Computing Engineering Department}}
\email{m.linaresv@uniandes.edu.co}
\author{Denys Poshyvanyk}
\affiliation{\institution{College of William and Mary} \department{Department of Computer Science}}
\email{denys@cs.wm.edu}

\renewcommand{\shortauthors}{K. Moran, C. Bernal Cardenas, M. Linares Vasquez et al.}

\begin{abstract}

Mobile devices and platforms have become an established target for modern software developers due to performant hardware and a large and growing user base numbering in the billions.  Despite their popularity, the software development process for mobile apps comes with a set of unique, domain-specific challenges rooted in program comprehension.  Many of these challenges stem from developer difficulties in reasoning about different representations of a program, a phenomenon we define as a ``language dichotomy".  In this paper, we reflect upon the various language dichotomies that contribute to open problems in program comprehension and development for mobile apps.  Furthermore, to help guide the research community towards effective solutions for these problems, we provide a roadmap  of directions for future work.

\end{abstract}



\copyrightyear{2018} 
\acmYear{2018} 
\setcopyright{rightsretained} 
\acmConference[ICPC '18]{ICPC '18: 26th IEEE/ACM International Conference on Program Comprehension }{May 27--28, 2018}{Gothenburg, Sweden}
\acmBooktitle{ICPC '18: ICPC '18: 26th IEEE/ACM International Conference on Program Comprehension , May 27--28, 2018, Gothenburg, Sweden}\acmDOI{10.1145/3196321.3196322}
\acmISBN{978-1-4503-5714-2/18/05}

\maketitle

\section{Introduction}
\label{sec:introduction}

	Mobile computing has become a centerpiece of modern society.  Smartphones and tablets continue to evolve at a rapid pace and the computational prowess of these devices is approaching parity with laptop and desktop systems for high-end mobile hardware.  This facilitates new categories of engaging software that aim to improve the ease of use and utility of computing tasks. Additionally, commodity smartphones are ushering in a completely new population of users from developing markets, many of whom are using a computer and accessing the internet for the first time.  These factors, combined with the ease of distributing mobile apps on marketplaces like Apple's App Store \cite{apple-app-store} or Google Play \cite{google-play} have made the development of mobile software a major focus of engineers around the world. In fact, according to Stack Overflow's 2018 survey of over 100,000 developers \cite{so-survey}, nearly a quarter of respondents identified themselves as mobile developers.  

	While the importance and prevalence of mobile in the modern software development ecosystem is clear, many of the unique attributes that make mobile platforms attractive to both developers and users contribute a varied set of challenges that serve as obstacles to producing high-quality software.  For example, while rich platform APIs facilitate development of advanced features, the change-prone nature of these APIs can adversely affect the quality of the apps they support \cite{Linares-Vasquez:FSE'13,Bavota:TSE15}. Another example of a mobile specific challenge relates to the touch-based, event driven nature of mobile apps.  Because the core functionality of many apps is driven mainly by the user interface, testing is typically performed at the GUI-level.  However, manual GUI-testing is a time-consuming task and developers need automated support to help reduce testing costs \cite{Linares-Vasquez:ICSME'17a,Linares-Vasquez:ICSME'17}.  While a sizable amount of work has been conducted to help automate mobile testing \cite{Choudhary:ASE'15}, many developers find that these approaches do not meet their needs \cite{Linares-Vasquez:ICSME'17a}.  

	When examining the current challenges that exist in mobile software development, maintenance, and testing one can observe a common thread weaved throughout these problems, contributing to a fabric of interconnected difficulties.  Incidentally, this thread is not something specific to mobile development, but rather stems from a fundamental trait of computer science more generally, namely \textit{abstraction}.  In their text \textit{``Foundations of Computer Science"} Aho and Ullman state that \textit{``fundamentally, computer science is a science of abstraction -- creating the right model for thinking about a problem and devising the appropriate mechanizable techniques to solve it."} Indeed, abstraction is a powerful concept in the engineering of software, allowing developers to design and implement complex programs.  However, there is also an associated cost that manifests itself when engineers must reason across multiple layers of abstraction.  In the domain of mobile development, abstractions contribute to and underlie many of the unique challenges experienced by developers.

	In particular, foundational abstractions between \textit{languages} prove to be particularly troublesome.  Here when we refer to the notion of a language we are not targeting programming languages specifically, but rather the broader definition of language as \textit{a medium by which an idea or information is conveyed}.  In this sense, there are several different languages, or modalities, of information that developers must navigate during the software development process for mobile applications, including natural language and code, just to name a few.  In essence, the bridging of the knowledge gap between these information modalities constitutes a set of principal challenges in \textit{program comprehension} for mobile apps.

	Specific development challenges can be viewed as arising from difficulties navigating different pairs of language types. For instance, when considering challenges related to change-prone APIs, developers must reason between program representations related to natural language and code, interpreting changes delineated in API documentation and how they may affect the use of those APIs in a particular app.  In GUI-based testing of mobile apps, developers and testers must reason between several different juxtaposed information modalities, including code and pixel-based representations of the app via the GUI.  In this paper we refer to these pairs of contrasting information modalities as \textit{language dichotomies}.  Developing solutions to help developers more effectively reason between various language dichotomies will help facilitate the resolution of many mobile development challenges.

	In this paper, we offer a brief introduction to mobile development paradigms (Section \ref{sec:mobile-background}), survey the major categories of research conducted to date on mobile software engineering (Section \ref{sec:related-work}), examine open challenges through the lens of language dichotomies (Section \ref{sec:challenges}), and outline a roadmap of potential future work aimed at aiding mobile developers in effectively navigating these dichotomies (Section \ref{sec:research-agenda}). It should be noted that this paper is by no means meant to be an exhaustive guide to the past research, processes or challenges related to developing mobile apps, but rather to prime the reader to think critically about the future research trends on the topic.  We hope that by examining key existing program comprehension problems related to mobile development from the viewpoint of language dichotomies, we can spur new, creative directions of work aimed at helping to solve these fundamental problems, which will in turn result in new processes and techniques for automating and facilitating software engineering for mobile apps.

\section{A Brief Introduction to Mobile Software Development}
\label{sec:mobile-background}

	In this section, we provide a brief overview of mobile development paradigms, as well as some of the attributes that make the mobile development process unique. Mobile applications are typically developed on top of an existing \textit{mobile platform}.  These platforms consist of several different parts and these parts can vary between platforms, however at a minimum usually include: (i) a kernel and an operating system (OS) that runs on mobile hardware such as a smartphone, (ii) an application framework consisting of a set of platform specific APIs and libraries, and (iii) a set of tools and software to aid in developing apps, including IDEs or user interface builders. Mobile apps are typically written using a target programming language supported for a particular platform (e.g., Java and Kotlin for Android, and Objective-C and Swift for iOS), in combination with the APIs from the platform's application framework. There are a shrinking set of platforms upon which developers can create and publish their apps. These platforms include Android, iOS, BlackBerry 10\footnote{\label{note1}Support will end at the end of 2019}, Firefox OS, Ubuntu Touch, and Windows 10 Mobile\textsuperscript{\ref{note1}}. However, currently Android and iOS comprise the majority of the market, accounting for 87.7\% and 12.1\% of the market share respectively for the $2^{\text{nd}}$ quarter of 2017\cite{statista-mobile-market-share-q2-2017}.  

\subsection{Unique Aspects of the Mobile Development Process}

\subsubsection{Platform Evolution and Instability}

Generally, the software development lifecycle typically follows a cyclic set of activities that include (i) requirement engineering, (ii) design, (iii) development, (iv) testing, and (v) maintenance.  Modern agile development practices typically iterate quickly through these activities with the goal of delivering working software in a continuous manner where features are added and bugs are fixed during each iterative development cycle. However, the rapid evolution of mobile platforms shapes the mobile development process in unique ways. As mobile hardware evolves, platforms evolve to keep pace with technological advancements, and new more convenient software features and capabilities are included with each iteration.  For instance, Android has had over 15 major version releases since its inception in 2008 that have dramatically reshaped the underlying platform APIs~\cite{McDonnell:ICSM'13}, leading to support for advanced features such as Augmented Reality (AR). This iterative process puts immense pressure on developers to evolve their apps with the mobile hardware and platforms to satisfy the expectations of users that their apps take advantage of the latest features~\cite{Jones:2014,Hu:EuroSys'14}.  This pressure leads to accelerated development cycles with a focus on adapting to changes in platform APIs.  Adapting to these changes can be difficult and may adversely affect app quality \cite{Linares-Vasquez:FSE'13,Bavota:TSE15}; because developers must cope with adding additional app functionality based on new platform features, or on fixing bugs that arise due to changes in APIs currently used in an app.  This may detract from time that could be spent on other activities such as fixing general regressions, refactoring, or improving the performance of an app, while also leading to undue technical debt. Thus, platform evolution has a clear affect on mobile development.

\subsubsection{GUI-Centric, Event Driven Applications}

	Perhaps one of the most important features of mobile devices is the ease of use provided by high-fidelity, touch-enabled displays.  Users primarily interact with their smartphones, tablets, and wearable devices and by extension the apps that run on these devices, through a touchscreen interface.  This means that mobile apps are centered around the graphical user interface, and are driven by touch events on this interface.  While other types of apps such as web apps, are also heavily event-driven, the unique touch based gestures and interactivity provided by mobile apps help to shape the software design, development and testing processes in unique ways.  For example, the user interface (UI) and user experience in mobile apps must be well-designed for an app to be successful in highly competitive marketplaces. As such design and development tools for constructing UIs are a core part of IDEs catering to mobile developers such as Xcode and Android Studio. Developers must constantly be aware of how application is connected to and impacts the GUI of their apps.

	The event-driven nature of mobile apps also impacts testing.  While developers can test small pieces of their code using practices such as unit testing, ultimately testing must be done through the GUI.  Manually testing applications is a time consuming practice that is fundamentally at odds with the rapid pace of mobile development practices.  Thus, mobile developers and testers will often utilize automation frameworks that either allow for reusable or fully automated test input generation.

\subsubsection{App Marketplaces}

	The primary (and some cases only) method of distribution for mobile apps is through ``app marketplaces" such as Google Play or Apple's App Store.  These digital storefronts are unique to mobile applications, in that they provide users with easy access to purchase, download, and update apps, while providing mechanisms for users to review apps and provide feedback to developers.  In recent years, these marketplaces have become increasingly competitive as the number of available apps numbers in the millions.  App marketplaces incentivize developers to ensure their apps are of the highest possible quality, and to take into account the feedback of users.  Developers need to ensure the quality of their apps by adhering to proper platform design principles and performing extensive testing, or risk being passed over for competitors.  Likewise developers need to react to feedback communicated through user reviews by gathering an updating requirements and updating their app's implementation.

\subsubsection{Market, Device, and Platform Fragmentation}

	The large and growing user base of smartphones and tablets is one of the most alluring aspects for many developers and companies hoping to reach users.  Unfortunately, targeting these users can be difficult due to multiple levels of fragmentation.  The first level of fragmentation is at the market-level, which is currently dominated by Android and iOS.  Thus, developers hoping to reach the maximum number of users must target \textit{both} of these platforms.  Second, there is fragmentation at the device level~\cite{Han:WCRE'12}, as there is a large and growing number of hardware options for consumers to choose from with more devices being introduced each year. Finally, there is platform fragmentation, as users on the same mobile platform may be running different versions of mobile OSes.  For instance, the latest version of iOS,  iOS11, is currently running on 65\% of devices whereas iOS10 currently encompasses 28\% of the install base~\cite{appstore-dashboard}.  However, in Android fragmentation is more severe where the two latest versions of Google's OS, Android 8 and 7, make up only 1.1\% and 28.5\% of the Android install base respectively.  In order to create effective apps, developers must ensure that their applications function properly across a wide of combinations of different platforms, devices, and platform versions.  This can make the process of developing and testing mobile apps challenging, as developers need to maintain concurrent codebases and test across a dizzying array of device and platform version configurations.

	Naturally, these difficulties have led to creation of platform-independent development tools such as Xamarin~\cite{xamarin}, where a single codebase can be compiled to multiple platforms, eliminating the need for parallel codebases.  Alternatively, there exist tools and frameworks like Ionic~\cite{ionic} for creating \textit{hybrid applications} which use a combination of web technologies that interface with underlying platform APIs. In addition to hybrid applications, another framework created by Facebook called React Native~\cite{react-native} facilitates the development of native mobile apps using javascript and React. Applications built using react native are fully native to the target platform, the framework simply assembles the native code according to the javascript written by a developer. All of these approaches can help ease the burden of fragmentation when creating mobile apps.  However, multi-platform development solutions come with their own set of compromises.  For instance, hybrid apps are known to suffer from performance issues in terms of user interface interactivity, which can frustrate users.  Furthermore, frameworks like Xamarin or React Native require their own learning curve, and developers are highly dependent upon the multi-platform framework keeping up with the latest features of modern mobile platforms.

\section{The State of Research in Mobile Software Development}
\label{sec:related-work}

	This section presents an overview of research related to software development in mobile ecosystems. We have segmented the current landscape of related work into seven major topics.  Note that the purpose of this section is to provide the reader with a primer on general research areas related to software development for mobile apps, we leave an in-depth systematic review as future work.

\subsection{App Store Analysis}

	App stores provide valuable information for users and developers. From user reviews to install base information, work on applying ``app store analytics" to help aid in the development process for mobile apps has seen great interest in recent years.  Recent work by Martin \etal~\cite{Martin:TSE'17} surveyed papers considering any type of technical or non-technical information from mobile markets. The authors categorized the papers into 7 different categories representing the underlying goal of empirical studies or new approaches for aiding the development process. The first of these is \textbf{API analysis} which constitutes papers that examine API usages in mobile apps. The second category, \textbf{feature analysis},  represents papers that extract and model both technical and non-technical information extracted from app stores. The third category, \textbf{release engineering}, analyzes release data and how this data can be used to help guide developers toward more effective release engineering. Fourth, \textbf{review analysis} considers all papers that analyze user reviews to extract information with the intention of using it to augment different parts of the mobile development process. App store analyses have also been conducted in relation to \textbf{security}, and this category describes papers that investigate how information from app stores can aid in security and the identification of  malware, faults, permissions, plagiarism, vulnerabilities, and privacy concerns on app stores. The sixth category, \textbf{store ecosystem}, includes papers analyzing the differences between app marketplaces. Last but not least is \textbf{size and effort prediction} which describes approaches that predict the effort or size of the functionalities. Recently there has been work on integrating information from user reviews to help aid in the testing process for mobile apps \cite{Grano:SANER'18}.

\subsection{App Security Analysis}

Mobile markets perform internal validations of apps to minimize the proliferation of malicious software and protect users privacy. In addition to the measures taken by application marketplaces, researchers have been actively engaged in using program analysis techniques to design new approaches for detecting malicious apps, analyzing security properties of applications, and assisting developers in creating apps with strong security principles. Sadeghi \etal~\cite{Sadeghi:TSE'17} performed a systematic literature review resulting in a taxonomy of research topics based on several complimentary dimensions that include the positioning of proposed approaches (\eg what problems are they trying to solve?), the characteristics of the approach (\eg how do they solve the purported problem?),  and finally the assessment of the approach (\eg How was it evaluated).  While we will refer readers to the full survey for more details, we examine the positioning of the examined papers here to provide an overview of the active research topics. The authors found the three main analysis objectives dominated the examined research, including \textbf{malware detection}, \textbf{vulnerability detection}, and \textbf{gray-ware detection}. Of these analysis objectives the examined papers targeted three major types of security threats, \textbf{spoofing}, \textbf{elevation of privilege}, and \textbf{information disclosure}.  These approaches used a variety of underlying program analysis techniques utilizing both static and dynamic information.

\vspace{-0.5em}
\subsection{Mobile Testing}

	Quality assurance is an important metric to be maintained in software development. This attribute is particularly important for mobile applications that will compete on fiercely competitive app marketplaces.  Performing effective testing is one of the best ways to ensure the quality of software produced and this topic has seen great interest from the software engineering research community. The largest area of work is focused on automated test input generation for mobile apps, and research in this area can be generally grouped into three categories~\cite{Choudhary:ASE'15}. The first category is, \textbf{random-based} input generation that randomly selects input events from a set of potential candidates\cite{Machiry:FSE'13,android-monkey,intent-fuzzer,Sasnauskas:WODA14,Ye:MoMM'13}. These random-based techniques may rely on a purely random event selection or generation function, or may bias the random selection based on the history of events with the aim of more effectively exploring an app under test.  The next type of approach, \textbf{systematic-based} input generation, follows a structured or well defined strategy for generating input events based upon a pre-defined heuristic for interacting with observable GUI-elements in an application. ~\cite{Azim:OOPSLA'13,Anand:FSE'12,Amalfitano:ASE'12,Moran:ICST'16,Gu:ICSME'17}. Finally, \textbf{model-based} input generation strategies create a model-based representation of a an application, which is then used to generate input events with one according to one of several goals such as uncovering crashes or covering the maximum number of program statements~\cite{Amalfitano:ASE'12,Yang:FASE'13,Azim:OOPSLA'13,Choi:OOPSLA'13,Hao:MobiSys'14,Zaeem:ICST'14,Linares-Vasquez:MSR'15,Mao:ASE'17,Mao:ISSTA'16}.

In addition to these strategies, there has also been work done on record and replay tools that allow developers to easily record GUI-level testing scenarios and replay them later as a form of regression testing~\cite{Gomez:ICSE'13,Hu:OOPSLA2015,Fazzini:ICST'17,Moran:MobileSoft'17}. Evaluating the efficacy of an automated test input generation technique can be challenging, as the practical utility of test adequacy criteria such as method or statement-level code coverage have come under scrutiny by the software testing research community. One potential alternative to these more traditional adequacy criteria is known as \textit{mutation analysis}.  This process purposefully injects faults into a software project and measures a test suite's ability to detect these faults.  However, for such a process to be effective, the fault injection techniques must seed faults representative of real errors that are likely to occur for a given software domain.  Thus, recent work has attempted to contextualize mutation testing for mobile apps, focusing on both functional and non-functional software quality attributes~\cite{Linares-Vasquez:FSE'17,Moran:ICSE'18-2,Deng:ICSTW'15,Deng:IST'17}.

\subsection{Building Effective User Interfaces}

	Creating effective UIs for mobile applications is a long and often tedious process that begins with UI mock-ups created by designers which are then given to development team to transfer these mock-ups into code that can be interpreted by mobile platforms~\cite{Moran:ICSE'18}. However, translating a mock-up of user interface into code can be a difficult undertaking.  Because developers can introduce errors when implementing the intended design of a mobile UI, there is a need for validation approaches to ensure the proper quality of mobile GUIs, and recent research has helped to enable such approaches. Joorabchi \etal~\cite{Joorabchi:ISSRE'15} presented an approach that validates the consistency between apps that are multi-platform, whereas Moran \etal~\cite{Moran:ICSE'18} focus on automatically reporting instances where the implementation of an Android GUI violates it's intended design specifications in an industrial context. Similarly, Fazzini \etal~\cite{Fazzini:ASE'17} conducted work that focuses on GUI validation in the context of comparing the behavior of the same app across platforms. 

	In addition to these approaches, there is a growing body of work that aims to automate the process of implementing a GUI from a mock-up outright, as any automation that can be introduced into the process can dramatically increase the efficiency of the overall mobile development process. REMAUI~\cite{Nguyen:ASE'15b} is a tool that aims at reverse engineering mobile interfaces by leveraging computer vision techniques. However, this only supports the generation for two types of UI components (text and images). Beltramelli \etal~\cite{Beltramelli:arXiv17} proposed and approach based on an encoder/decoder model for translating images into a Domain Specific Language (DSL) which can then be converted into code. However, this approach was only tested on a small set of synthetic apps, and has yet to be proven on real applications.  ReDraw~\cite{Moran:ArX'18}, aims to overcome the limitations of both pix2code and REMAUI, by mining GUI-related information from app stores and using machine learning approaches to help build a realistic GUI-hierarchy which can be automatically translated into code.

\subsection{Static Program Analysis for Mobile Apps}

Li \etal~\cite{Li:IST'17} conducted a systematic literature review taxonomizing work done on static analysis for Android applications. This review found that the most popular aims of static analysis tools for Android were: (i) \textbf{data leak detection}, (ii) \textbf{vulnerability detection}, (iii), \textbf{permission analysis}, (iv) \textbf{energy analysis}, and (v) \textbf{clone detection}. Moreover, the Smali and Jimple intermediate representations were the most widely used program representations. Regarding the analysis methods used by these techniques, there have been approaches that use abstract interpretation, taint analysis, symbolic execution, program slicing, code instrumentation, and type/model checking. Several of these approaches target Android specific program constructs including the component lifecycle, UI callbacks, entry points, inter-component communication, inter-app communication, and XML layout.

\subsection{Energy \& Performance Analysis}

	Nearly all mobile devices operate in a resource constrained context and draw power from a battery.  Thus, the non-functional attributes of mobile software, such as performance and energy efficiency, have been a popular topic of study among researchers. This body of work is comprised of empirical studies that study these topics in depth and approaches for improving non-functional aspects of mobile apps during the development process. This work can generally be classified into the following categories: (i) \textbf{estimation and prediction of energy consumption}~\cite{DiNucci:ICSE'17C,DiNucci:SANER'17,Chowdhury:MSR'16,Romansky:ICSME'17} (ii) \textbf{energy consumption of GUIs} \cite{Dong:TMC'12,Dong:TMC'12-2,Li:ICSE'14,Wan:ICST'15,Stanley:CC'16,Agolli:MobileSoft'17,Linares-Vasquez:FSE'15,Linares-Vasquez:ICSE-C'17} (iii) \textbf{energy bugs and hot spots}~\cite{Pathak:Hotnets'11,Pathak:MobiSys'12,Pathak:EuroSys'12,Linares-Vasquez:MSR'14,Li:ICSME'14,Li:ISSTA'13,Liu:TSE'14,Li:DeMobile'15,Li:ICSE'16,Wu:CC'16,Rasmussen:GREENS'14,Gui:ICSE'15} (iv) \textbf{energy consumption considering other factors} such as memory, obfuscation, CPU usage among others ~\cite{Sahin:JSEP'16,Jabbarvand:GREENS'15,Saborino:CoRR'17}.

	Other approaches have focused on measuring the performance on mobile apps. For instances, Linares-V\'{a}squez \etal~\cite{Linares-Vasquez:ICSME'15} surveyed developers to determine best practices and tools that could be used to avoid performance bottlenecks. Similarly, Lin \etal~\cite{Lin:ASE'15b} implemented a tool to refactor \texttt{AsynTask} to avoid memory leaks and reduce energy consumption. Moreover, Linares-V\'{a}squez \etal~\cite{Linares-Vasquez:JSS'17} studied micro-optimizations opportunities, reductions on memory and CPU performance, and developers' practices on micro-optimizations on Android.

\subsection{Mobile Fragmentation}

As overviewed earlier (Section~\ref{sec:mobile-background}) fragmentation is a well known problem by developers of mobile applications. Han \etal~\cite{Han:WCRE'12} give an excellent overview on a topic-model based analysis evidencing the lack of portability and fragmentation considering multiple vendors. Moreover, McDonnell \etal~\cite{McDonnell:ICSM'13} analyzed change prone Android APIs and examined how quickly these changes are adopted in apps. The results of this study demonstrated slow adoption in several cases. Other approaches have focused on providing strategies to prioritize the devices upon which a developer should focus app testing~\cite{Khalid:FSE'14, Lu:ICSE'16a}. In contrast to these approaches, Wei \etal~\cite{Wei:ASE'16} focused their attention on detecting and understanding compatibility issues at code level. Finally, Linares-V\'{a}squez \etal \cite{Linares-Vasquez:FSE'13} and Bavota \etal \cite{Bavota:TSE15} analyzed the impact of rapid changes in the Android platform to application ratings on Google Play.

\section{Challenges in Program Comprehension for Mobile Apps}
\label{sec:challenges}

	There is no doubt that significant progress on understanding and improving the mobile development process has been made due to the large and growing body of research from the software engineering community.  However, there still exist sizable challenges that must be properly investigated and solved in future work \cite{Muccini:AST'12}.  As stated at the outset of this paper, many of these open challenges share a common trait; they arise due to various \textit{language dichotomies} that developers must reason about in order to build, test, and maintain successful apps.  More specifically, a language dichotomy can be defined as \textit{a difficulty in program comprehension resulting from reasoning about different representations or modalities of information that describe a program.}  In the domain of mobile applications there are several language dichotomies that contribute to a varied set of problems.  In this section we will examine the problems resulting from dichotomies involving four major modalities of information:

\begin{enumerate}
	\item{\textbf{\textit{Natural Language}:} This modality represents languages that humans typically use to convey ideas or information to one another, such as English.}
	\item{\textbf{\textit{Code}:} This modality represents the languages that humans utilize to construct a program, such as Java or Swift.}
	\item{\textbf{\textit{Graphical User Interfaces (GUIs)}:} Much of today's user facing software is graphical, and mobile apps are no exception.  This information modality is highly visual, consisting of pixel-based representations of a program typically comprised of a logical set of building blocks often referred to as GUI-widgets or GUI-components.}
	\item{\textbf{\textit{Dynamic Program Event Sequences}:}  As a mobile application is executed, the series of inputs, events, and program responses to these events represents a rich modality of information that describes program behavior.}
\end{enumerate}

	Each of the representations described above have their own powerful uses, often serving to facilitate program abstractions.  For example, a GUI is an extremely powerful abstraction of program code that allows for seamless interaction and use of features.  However, for a developer, it is often critical to effectively understand and navigate how information represented in one modality translates to another.  This is, at its core, a \textit{program comprehension} task.  For instance, a developer must reason about how different parts of the GUI correspond to different sections of code in a mobile app.  However, bridging this gap between representations can be an arduous task, and thus underlies many open problems in mobile software development.

	In this section we overview five language dichotomies consisting of the information modalities listed above and the mobile development problems that stem from them.  Note that this is not meant to be an exhaustive list of language dichotomies or problems, but rather a curated list based upon our research observations of the past several years.  We encourage readers to seek out and define new problems which we may not have discussed in detail.

\subsection{Natural Language vs. Code}
\label{subec:nlvscode}

Perhaps one of the most well-known language dichotomies for developers is that between natural language and code.  This dichotomy often surfaces when software requirements or specifications are stipulated in natural language before being implemented in code.  In this instance, developers must bridge this language gap and reason about the code-based representation of the information encoded into the natural language.  In the context of mobile development, reasoning about this dichotomy is exacerbated. This is not due to the size or relative complexity of mobile apps, but instead to their event driven nature and varying contextual states.  Tracing features to code constructs in a mobile app can be difficult due to the disconnect between event-handlers, platform APIs, functional code, context (\eg network and sensor data) and connection to the GUI-code.  Thus, implementing and reasoning about features represented in natural language can quickly become an intensive task. In our experience, this dichotomy has contributed to two important open problems in mobile app development.  

\subsubsection{Feature Location and Traceability}

	Feature Location has been defined as \textit{``the activity of identifying an initial location in the source code that implements functionality in a software system"}~\cite{Dit:EMSE'13}.  Feature location is an important program comprehension task in software development and maintenance, as it is one of the most frequent developer activities.  A wealth of research has been conducted related to feature location techniques, however, few of these techniques have been contextualized and applied to the domain of mobile applications.  The most closely related work on feature location for mobile apps stems from work conducted by Palomba \etal that recommends and localizes code changes based on information from user reviews \cite{Palomba:ICSE'17}.  However, little work has been conducted that attempts to link requirements or features, stipulated in natural language, to code-related program constructs for the purpose of supporting developer comprehension.

	Feature Location is particularly relevant in the context of mobile software due to constant pressure for developers for frequent releases to keep up with the rapid evolution of mobile platforms and hardware~\cite{Hu:EuroSys'14,Jones:2014,Linares-Vasquez:FSE'13,Bavota:TSE15}. Because developers are changing the source code often, they will have to continually locate and understand features in the source code.  Due to the event driven nature of mobile apps, developers need adequate support for this intellectually intensive task.  Such support for developers has the potential to greatly increase productivity and improve the efficiency and effectiveness of the software maintenance and evolution processes.

	Software traceability generally describes the process of establishing relationships between software requirements and code. While there has been a large body of work devoted to enabling effective software traceability, few of these techniques have specifically targeted the domain of mobile applications.  Traceability is important during the mobile development process for developers to ensure that requirements are properly implemented and tested in the source code.  However, mobile apps present a set of unique challenges for traditional software traceability approaches.  For instance, mobile applications have access to sensitive user information that can be collected from a diverse set of sensors such as location, or user audio.  Most popular mobile platforms, including iOS and Android, implement a permission system that allows a user to grant access potentially sensitive user information or hardware sensors. Given the importance of these permission systems in user privacy, they must be effectively taken into account by traceability approaches, and security and privacy related requirements should consider the permissions systems and other security measures implemented in code.  This requires reasoning between natural language descriptions of permissions and security principles while linking this information to relevant areas of code.  Another unique attribute of mobile applications that must be taken into consideration is the heavily used set of platform APIs used to implement large amounts of the app functionality. Traceability approaches must be cognizant of the natural language documentation and API code to establish accurate trace links.

\subsubsection{Bug and Error Reporting}

	Bug and Error reporting is an important activity for any type of software system, and techniques for bug triaging~\cite{Shokripour:MSR'13,Naguib:MSR'13,Jeong:ESEC/FSE'09,44Kim:TOSE2013,Kim:DSN'11,Aranda:,Linares-Vasquez:a}, duplicate report detection~\cite{Joorabchi:MSR'14, Nguyen:ASE2012, Wang:ICSE'08, Guo:ICSE'10,  Zhou:CIKM'12, Gu:ICSE'10}, summarization~\cite{Mani:FSE'12,Bettenburg:MSR'08,Rastkar:ICSE'10,33Koru:IEEE2004,Weiss:MSR'07,Czarnecki:ICSM'12}, and reporting of in-field failures~\cite{Bell:ICSE'13, Jin:ISSTA'13, Zhou:ICSE'12, Clause:ICSE'07,Jin:ICSE'12,Kifetew:ICST'14,Cao:ASE'14} have been devised to help improve this process.  In the domain of mobile apps, the primary mechanism by which feedback and bug reports are communicated to developers is through user reviews on app stores.  These user reviews have been shown to be incredibly noisy \cite{Chen:ICSE'14a} and a large body of work has been dedicated to extracting effective information from these reviews and operationalizing it to help aid in software development and testing tasks.  While this research has proven to be valuable, little work has been conducted to help improve the relatively rudimentary mechanisms employed by App Stores to provide feedback.  

	At its core, the process of bug reporting and resolution requires bridging a knowledge gap between high-level program features (often described in natural language) and program information represented in code.  Our past work on the \fusion bug reporting system \cite{Moran:FSE'15,Moran:ICSE'16} aims to help bridge this gap by \textit{improving the underlying mechanism by which users report bugs}. Furthermore, our work on \crashscope has helped to automate the bug reporting process outright for program crashes. While this work showed that automating and reinventing the the bug reporting process has great promise, much more work needs to be done in bridging the language dichotomy that exists in bug reports.  This is particularly important for mobile apps, as their event driven nature and varying contextual states can contribute to bug reproduction scenarios that are difficult to stipulate in natural language, and thus may need more advanced reporting mechanisms.

\subsection{Code vs. Graphical User Interfaces}
\label{subec:codevsgui}

	As with most modern user-facing software, mobile applications are heavily centered around their graphical user interfaces (GUIs).  While GUIs may not be considered a traditional language or modality in which program information is encoded, they contain a wealth of practical data that can be used to help reason about software properties.  GUI information is intrinsically linked with an applications' higher level functional and non-functional features. Furthermore, the GUI specifications are typically stipulated in source code (\eg the \texttt{\small /res/layout/} folder of Android apps) and thus is inherently linked to code constructs. While mappings between program features and code exist, the ambiguities that exist between these representations can often be difficult to overcome.  In modern mobile development, GUIs must be dynamic and reactive to adapt to an increasing number of hardware configurations and screen technologies.  However, this means that GUIs are often adjusted dynamically at runtime, decoupling runtime GUIs from code specifications.  Furthermore, most modern mobile apps also rely upon network connectivity features to pull information from the internet, and thus a majority of the content displayed by a mobile app's GUI is dynamic and directly stipulated in code.  These are just two examples of existing ambiguities that complicate the language dichotomy between GUIs and code. GUI-related information is often underutilized in research related to solving practical program comprehension problems, and we highlight two instances of open challenges in mobile apps that could be mitigated by working to close the abstraction gap between code and user interfaces.

\subsubsection{Visualizing the Affects of Code Changes on the GUI}

	Due to the GUI-centric nature of mobile apps, developers must constantly reason about how their code affects and is connected to the GUI.  However, this process can be incredibly tedious, as developers must switch contexts between code, visual representations, and markup-like code that stipulates the visual properties of the user interface.  Currently, IDEs for the two most popular platforms provide support within the IDE for building GUIs and visualizing the layout of an application during development \cite{xcode-interface,android-interface}.  However, such features are typically limited to illustrating the properties of the GUI-related code only (\eg xml markup files in Android), or to event handlers (\eg XCode).  Developers need further support for visualizing how logical code is linked to different parts of the GUI during the mobile application development process. 

\subsubsection{Ensuring the Proper Implementation of GUIs from Design Specifications}

	The UI/UX design for mobile applications is becoming increasingly important in competitive app marketplaces.  As many applications target similar core functionality (\eg weather apps, task managers) they must differentiate themselves with attractive user interfaces and intuitive user experiences.  As such, many companies employ a dedicated team of designers with domain expertise in creating visually striking and easy to use GUIs.  Even independent developers not part of a large corporation will often create user interface mockups to prove out design ideas and test UI concepts before committing to implementing them. In both cases, these professionals will often use design software such as Sketch \cite{sketch} or Photoshop \cite{photoshop}, generally due to the flexibility offered by these tools. Once these mock-ups have been created, they must be implemented in code by developers, a process that has been shown to be time consuming and error-prone \cite{Tucker:CSH04,Myers:CHD94,Nguyen:ASE'15,Lelli:ICST'15, Moran:ICSE'18}. 

	Developers and designers need support throughout this process in order to enable effective prototyping of mobile application user interfaces, which involves bridging an abstraction gap between graphical and code-based representations of a program.  Initial work on this problem has been done from two viewpoints: (i) automatically reporting instances where an implementation of a GUI does not match its intended design specifications \cite{Moran:ICSE'18,Mahajan:ICSE'18,Fazzini:ASE'17}, and (ii) automating the process of prototyping a GUI from a mock-up \cite{Nguyen:ASE'15b,Moran:ArX'18,Beltramelli:arXiv17}.  However, there are still several problems to be solved to aid in facilitating and automating the process of implementing a GUI, and the underlying app functionality, from a mock-up or series of mock-ups.  For instance, little work has been conducted in automatically implementing transitions between related screens, or generating code related to the underlying functionality of different GUI-components.

\subsubsection{Augmented Reality}

	Smartphones have evolved to become incredibly capable devices, with computational prowess that is beginning to rival more traditional laptop computers. This combined with the rapid advancement of many sensors, most notably cameras, has ushered in new use cases that take unique advantage of increasingly capable hardware.  Perhaps the most notable of these new use cases is commonly referred to as \textit{Augmented Reality} (AR).  AR applications typically aim to enhance or ``augment" a users physical world by simulating projections of useful information or graphics into the real world using a camera and a display.  This can facilitate, for example, digital projections of furniture onto a video stream of a users home or apartment using a smartphone camera and display.  Apple and Google have both recently supported this technology with the release of ArKit for iOS~\cite{arkit} and ARCore for Android~\cite{arcore}.  While this new category of applications brings with it exciting new use cases, the development challenges of such types of applications have yet to be explored thoroughly.  Surely applications implementing such unique features will offer unique challenges from the point of view of user interface design and testing, however, researchers need to better understand such challenges and develop techniques and tools to help facilitate the creation of AR apps.

\subsection{Natural Language vs. Graphical User Interfaces}
\label{subec:nlvsgui}

	While GUIs are inherently interconnected with code, they also form dichotomy with natural language.  Since much of an app's functionality is associated with the actions a user can perform on the GUI, there is clear link between natural language describing app features and GUI-based representations of an app.  Bridging this gap is a necessary task for developers, and there has been little work to help facilitate this process.

\subsubsection{Use Case-Based Testing}

	One area that could greatly benefit from bridging the abstraction gap between natural language and GUIs is automated testing.  Due to the centrality of the GUI in exposing most program functionality for mobile apps, testing is typically conducted at the GUI level.  However, mobile developers have specific testing needs, and while automated approaches for finding crashes exhibit some utility, many mobile developers prefer to organize their tests around use-cases~\cite{Linares-Vasquez:ICSME'17a}. However, automating test case generation around use cases can be difficult, even if the use-cases are stipulated in natural language. This difficulty stems from the fact that the test generation approach must effectively navigate the language dichotomy between features and use cases stipulated in natural language, and information displayed by an application's GUI to generate a sensible sequence of test input events.  In the absence of existing natural language use cases, an automated approach would have to infer, \textit{online}, the use cases of the app in natural language so that they could be documented and effectively understood by a developer. Initial work on modeling app events have been conducted through the \MonkeyLab project \cite{Linares-Vasquez:MSR'15}, however, such work needs to be taken further in order to enable practical use-case based testing for developers.

\subsubsection{Protecting User Privacy in Mobile Apps}

	In the last few years, privacy become an even more critical component of the software development process as users store more sensitive information in digital spaces than ever before.  Mobile developers also need to be continuously aware of the security implications of the software they create, as the capabilities of mobile phones can enable the collection of intimate, sensitive user data such as user location and audio/visual recordings.  A large component of the security and privacy of mobile apps involves informing users how their data is being utilized by software.  However, in practice this can be difficult or cumbersome for developers to implement, as it involves reasoning between natural language descriptions of privacy information and effective incorporation into the GUI.  Further work needs to be conducted to better support and automate the process of informing the user about the use of security or privacy related features of mobile apps.

\subsection{Event Sequences vs. Natural Language \& GUIs}
\label{subec:etvsnlvscode}

	Modeling the behavior of mobile applications has been a popular topic related to automated testing approaches for mobile apps~\cite{Linares-Vasquez:MSR'15,Mao:ASE'17}.  Many of these approaches use event sequences traces to help model application behavior and generate more useful testing sequences.  However, the representative power of these models suffers due to language dichotomies that exist between the event traces and code as well as between event traces and natural language. For instance, relating event sequences to natural language descriptions of features or bugs could help guide automated test generation towards certain testing goals more effectively.

\subsubsection{Cross Platform and Cross Device Testing}

	One well understood problem in mobile development, and more specifically for Android development, is that of device and platform fragmentation~\cite{Han:WCRE'12}.  Due to the plethora of devices running various versions of underlying platform software, developing an mobile application that functions seamlessly across all of these platforms is a major challenge for developers.  One of the biggest challenges related to the development process is testing an application across a large combinatorial matrix of physical devices and hardware versions. Ideally, developers could write a single test case and have this test case effectively operate across multiple devices, platforms (\eg iOS and Android), and platform versions (\eg iOS 10 vs. iOS 11). While some existing work has been done toward enabling such testing approaches~\cite{Fazzini:ICST'17}, this remains an open problem and general pain point for mobile development and testing. To help mitigate this problem, event sequences need to be translated across applications with varying differences automatically, which involves abstracting or modeling the event sequences across differing GUIs, and perhaps relating these changes to code differences as well.

\subsubsection{Understanding the Affect of Software Evolution on Use Cases}

		Due to the highly iterative nature of underlying platform APIs and hardware, mobile applications tend to evolve at a rapid pace. However, timelines for app releases are tight and often developers do not have sufficient time or resources to properly document all aspects of an application's evolution.  One such property of apps that can difficult to document are changes in the use cases, or changes to the event sequences required for a user to carry out existing use cases.  Properly documenting these software development artifacts carries implications for enabling effective testing, traceability, and feature location. Thus, this topic deserves ample attention from researchers.

\section{Future Trends in Program Comprehension for Mobile Apps}
\label{sec:research-agenda}

We expect future research in mobile software engineering to be driven by need to deal with language dichotomies and the aforementioned challenges. Thus, in this section we discuss likely future trends in mobile software engineering research that share a common goal of helping to solve language dichotomies that contribute to challenges in program comprehension.

\noindent \textbf{\textit{Natural Language vs. Code:}} While app marketplaces continue to be the preferred platform for app distribution, short release cycles will continue to burden mobile developers as they consistently attempt to appease the collective voice of users. Thus, the current and future mechanisms for gathering user feedback must be oriented to reduce the language gap between the changes that users request, and the incorporation of these "change requests" into codebases and tests. Automated linking of user reviews and bug reports to source code is a first step partially achieved by current research \cite{Palomba:ICSME'15,Palomba:ICSE'17,Palomba:JSS17}; next steps should be devoted to enable automated generation, prioritization, and execution of test cases but triggered by incoming user reviews and crashlytic data collected at run-time, without human intervention \cite{Linares-Vasquez:ICSME'17}. 

	However, the larger challenge here is related to understanding user needs that are expressed in very short snippets of text which may include very personal expressions, jargon, acronyms, or domain-specific language. One potential solution here is to move from text-based reviews/requests/reports to augmented representations that remove the ambiguities in natural language.  Some potential options for such representations might include on-device data collection, behavior-driven specifications, sketch-based reviews, or video-based bug reporting. Another potential solution might be to include advanced machine learning mechanisms that learn from user reviews and are able to extract high level concepts and relationships (\eg by using deep learning) that can be automatically translated into code or tests.

	Another developer need that is closely aligned with shorter DevOps cycles in mobile apps is automated source code generation assisted by high-level representations. The recent introduction of software architecture components in Android~\cite{android-architecture} makes it easier to create applications that are designed to follow well established patterns (Views, Controllers, ViewModels, DAOs, entities, etc); in the case of iOS, the usage of the MVC architectural pattern is well established. New techniques for automated code generation could leverage these architectural design patterns, in combination with models of code and natural language mined from software repositories to enable practical code generation.  Another challenge here is to automatically handle API breaking changes that can be difficult for developers to identify due to the continuous releases of new API versions (as in the case of Android), and also because current mechanisms for reporting changes in the APIs are detached from the app development process. Future work should examine better methodologies for incorporating information about API changes into the development workflow.

	In summary, enabling automated generation of source code that follows the architectural patterns proposed by each platform, and that is up-to-date with the latest API versions will be an important trend in coming years. Having such approaches/tools will help developers to be more focused on designing mobile apps with better UX/UI and less prone to issues imposed by the fragmentation.

\noindent \textbf{\textit{GUIs vs. Code and Natural Language:}} Designing for a multi-device experience (sequential and complementary) is becoming more common today as users demand more intricate integration of mobile apps/devices with different devices/appliances available across a range of different contexts (office, home, public transportation, etc.). For example, users may utilize an application across a smartphone, tablet, wearable device, and digital voice assistant. Enabling such multi-device experiences by default necessitates cross-platform applications. Current approaches for multi-device or multi-platform mobile app development and testing still leave much to be desired from a developers point of view, as nearly all current approaches come with undesirable trade-offs. For example, UI performance issues related to hybrid applications. However, this phenomenon represents a ripe research opportunity, in particular for dealing with language dichotomies between functional code, GUI-code, and pixel-based representations of GUIs. Developers desperately need models and frameworks that are able to express the interaction of apps across multiple-devices and multiple-platforms in such a manner that allows for designing-once-running-everywhere and designing-once-testing-everywhere.  Research should focus on converging upon such a solution, as this would help mitigate several key challenges in program comprehension for mobile apps.

\noindent \textbf{\textit{Event Traces vs. Code and Natural Language:}} As multi-platform and multi-device apps become a more necessary part of mobile development, it is important that event-sequences are properly modeled across different contexts.  This means that mobile developers and testers need a method of abstracting the individualized event-sequences that exist for a given platform or device, to a more general representation, linked to natural language descriptions, that are portable between devices and platforms.  This would allow for a unified understanding of high-level functional use cases across apps expressed in natural language, while having positive implications for test case generation and maintenance.  Researchers should examine new methods of modeling such relationships to help make such a unified representation of application events sequences a reality.  One potentially promising modeling technique might come by the way of emerging deep learning algorithms for machine translation.

 \noindent \textbf{\textit{All Dichotomies:}} Finally, we see the On-Demand Developer Documentation (OD3) paradigm \cite{Robillard:ICSME'17} as a vision supporting the goal of reducing the gaps in language dichotomies. OD3 systems could be used to generate documentation able to serve as the linking points between language GUI, code, and event sequences. Therefore, we support the OD3 vision, and encourage mobile software engineering researchers to propose systems that are aligned with goals set forth in OD3 and tailored to mobile development challenges.

\section{Conclusion}
\label{sec:conclusion}

	In this paper, we introduced the idea of a \textit{language dichotomy} as an abstraction gap between contrasting information modalities in software that contribute to challenges in program comprehension.  We then provided a brief summary of the unique aspects of the mobile development process, as well as the research that has been conducted to help understand issues and improve the process as a whole.  Using the notion of a language dichotomy as a guide, we examined several open challenges related to program comprehension during the development of mobile apps. Finally, we reviewed a potential research agenda aimed at overcoming the fundamental language dichotomies that contribute to a wide range of challenges in program comprehension for mobile apps, with the hope that researchers will use this as starting point for working towards bridging the gap between different information modalities of mobile software.

\begin{acks}
This work is supported in part by the NSF CCF-1218129, NSF CCF-1253837, and NSF CCF-1525902 grants.  Any opinions, findings, and conclusions expressed herein are the authors' and do not necessarily reflect those of the sponsors.
\end{acks}

\balance
\bibliographystyle{abbrv}

\bibliography{ms}

\begin{thebibliography}{100}

\bibitem{so-survey}
2018 stack overflow developer survey
  \url{https://insights.stackoverflow.com/survey/2018/}.

\bibitem{photoshop}
Adobe {{Photoshop}} {{http://www.photoshop.com}}.

\bibitem{android-architecture}
Android architecture components
  \url{https://developer.android.com/topic/libraries/architecture/index.html}.

\bibitem{android-interface}
Android studio layout editor
  \url{https://developer.android.com/studio/write/layout-editor.html}.

\bibitem{android-monkey}
Android {{UI}}/{{Application Exerciser Monkey}}
  {{http://developer.android.com/tools/help/monkey.html}}.

\bibitem{apple-app-store}
Apple {{App Store}} {{https://www.apple.com/ios/app-store/}}.

\bibitem{arcore}
Arcore \url{https://developers.google.com/ar/discover/}.

\bibitem{arkit}
Arkit \url{https://developer.apple.com/arkit/}.

\bibitem{google-play}
Google {{Play Store}} {{https://play.google.com/store?hl=en}}.

\bibitem{intent-fuzzer}
Intent {{Fuzzer}}
  {{https://www.isecpartners.com/tools/mobile-security/intent-fuzzer.aspx}}.

\bibitem{ionic}
ionic framework \url{https://ionicframework.com/}.

\bibitem{react-native}
React native \url{https://facebook.github.io/react-native/}.

\bibitem{sketch}
The {{Sketch Design Tool}} {{https://www.sketchapp.com}}.

\bibitem{statista-mobile-market-share-q2-2017}
{{Statista - Mobile Market Share}}
  {{https://www.statista.com/statistics/266136/global-market-share-held-by-smartphone-operating-systems/}}.

\bibitem{xamarin}
Xamarin {{Test Cloud}} {{https://www.xamarin.com}}.

\bibitem{xcode-interface}
Xcode interface builder
  \url{https://developer.apple.com/xcode/interface-builder/}.

\bibitem{Linares-Vasquez:JSS'17}
How developers micro-optimize android apps.
\newblock {\em J. Syst. Softw.}, 130(C):1--23, Aug. 2017.

\bibitem{Agolli:MobileSoft'17}
T.~Agolli, L.~Pollock, and J.~Clause.
\newblock Investigating decreasing energy usage in mobile apps via
  indistinguishable color changes.
\newblock In {\em 2017 IEEE/ACM 4th International Conference on Mobile Software
  Engineering and Systems}, MobileSoft'17, pages 30--34, May 2017.

\bibitem{Amalfitano:ASE'12}
D.~Amalfitano, A.~R. Fasolino, P.~Tramontana, S.~De~Carmine, and A.~M. Memon.
\newblock Using {{GUI Ripping}} for {{Automated Testing}} of {{Android
  Applications}}.
\newblock In {\em Proceedings of the 27th {{IEEE}}/{{ACM International
  Conference}} on {{Automated Software Engineering}}}, ASE'12, pages 258--261,
  Essen, Germany, 2012. {ACM}.

\bibitem{Anand:FSE'12}
S.~Anand, M.~Naik, M.~J. Harrold, and H.~Yang.
\newblock Automated {{Concolic Testing}} of {{Smartphone Apps}}.
\newblock In {\em Proceedings of the {{ACM SIGSOFT}} 20th {{International
  Symposium}} on the {{Foundations}} of {{Software Engineering}}}, FSE '12,
  pages 59:1--59:11, Cary, North Carolina, 2012. {ACM}.

\bibitem{appstore-dashboard}
{Apple}.
\newblock App {{Store}} - {{Support}}.
  {{https://developer.apple.com/support/app-store/}}.

\bibitem{Aranda:}
J.~Aranda and G.~Venolia.
\newblock The secret life of bugs: {{Going}} past the errors and omissions in
  software repositories.
\newblock In {\em 31st {{International Conference}} on {{Software
  Engineering}}, {{ICSE}} 2009, {{May}} 16-24, 2009, {{Vancouver}}, {{Canada}},
  {{Proceedings}}}, pages 298--308, 2009.

\bibitem{Azim:OOPSLA'13}
T.~Azim and I.~Neamtiu.
\newblock Targeted and {{Depth}}-first {{Exploration}} for {{Systematic
  Testing}} of {{Android Apps}}.
\newblock In {\em Proceedings of the 2013 {{ACM SIGPLAN International
  Conference}} on {{Object Oriented Programming Systems Languages}} \&\#38;
  {{Applications}}}, OOPSLA '13, pages 641--660, Indianapolis, Indiana, USA,
  2013. {ACM}.

\bibitem{Bavota:TSE15}
G.~Bavota, M.~Linares-V{\'a}squez, C.~Bernal-C{\'a}rdenas, M.~Di~Penta,
  R.~Oliveto, and D.~Poshyvanyk.
\newblock The {{Impact}} of {{API Change}}- and {{Fault}}-{{Proneness}} on the
  {{User Ratings}} of {{Android Apps}}.
\newblock {\em Software Engineering, IEEE Transactions on}, 41(4):384--407,
  Apr. 2015.

\bibitem{Bell:ICSE'13}
J.~Bell, N.~Sarda, and G.~Kaiser.
\newblock Chronicler: {{Lightweight Recording}} to {{Reproduce Field
  Failures}}.
\newblock In {\em Proceedings of the 2013 {{International Conference}} on
  {{Software Engineering}}}, ICSE'13, pages 362--371, San Francisco, CA, USA,
  2013. {IEEE Press}.

\bibitem{Beltramelli:arXiv17}
T.~Beltramelli.
\newblock Pix2code: {{Generating Code}} from a {{Graphical User Interface
  Screenshot}}.
\newblock {\em CoRR}, abs/1705.07962, 2017.

\bibitem{Bettenburg:MSR'08}
N.~Bettenburg, R.~Premraj, T.~Zimmermann, and S.~Kim.
\newblock Extracting {{Structural Information}} from {{Bug Reports}}.
\newblock In {\em Proceedings of the 2008 {{International Working Conference}}
  on {{Mining Software Repositories}}}, MSR '08, pages 27--30, Leipzig,
  Germany, 2008. {ACM}.

\bibitem{Cao:ASE'14}
Y.~Cao, H.~Zhang, and S.~Ding.
\newblock {{SymCrash}}: {{Selective Recording}} for {{Reproducing Crashes}}.
\newblock In {\em Proceedings of the 29th {{ACM}}/{{IEEE International
  Conference}} on {{Automated Software Engineering}}}, ASE '14, pages 791--802,
  Vasteras, Sweden, 2014. {ACM}.

\bibitem{Chen:ICSE'14a}
N.~Chen, J.~Lin, S.~C.~H. Hoi, X.~Xiao, and B.~Zhang.
\newblock {{AR}}-miner: {{Mining Informative Reviews}} for {{Developers}} from
  {{Mobile App Marketplace}}.
\newblock In {\em Proceedings of the 36th {{International Conference}} on
  {{Software Engineering}}}, ICSE'14, pages 767--778, Hyderabad, India, 2014.
  {ACM}.

\bibitem{Choi:OOPSLA'13}
W.~Choi, G.~Necula, and K.~Sen.
\newblock Guided {{GUI Testing}} of {{Android Apps}} with {{Minimal Restart}}
  and {{Approximate Learning}}.
\newblock In {\em Proceedings of the 2013 {{ACM SIGPLAN International
  Conference}} on {{Object Oriented Programming Systems Languages}} \&\#38;
  {{Applications}}}, OOPSLA '13, pages 623--640, Indianapolis, Indiana, USA,
  2013. {ACM}.

\bibitem{Choudhary:ASE'15}
S.~R. Choudhary, A.~Gorla, and A.~Orso.
\newblock Automated {{Test Input Generation}} for {{Android}}: {{Are We There
  Yet}}? ({{E}}).
\newblock In {\em 2015 30th {{IEEE}}/{{ACM International Conference}} on
  {{Automated Software Engineering}} ({{ASE}})}, ASE'15, pages 429--440, Nov.
  2015.
\newblock ISSN:.

\bibitem{Chowdhury:MSR'16}
S.~A. Chowdhury and A.~Hindle.
\newblock Characterizing {{Energy}}-aware {{Software Projects}}: {{Are They
  Different}}?
\newblock In {\em Proceedings of the 13th {{International Conference}} on
  {{Mining Software Repositories}}}, MSR '16, pages 508--511, Austin, Texas,
  2016. {ACM}.

\bibitem{Clause:ICSE'07}
J.~Clause and A.~Orso.
\newblock A {{Technique}} for {{Enabling}} and {{Supporting Debugging}} of
  {{Field Failures}}.
\newblock In {\em Proceedings of the 29th {{International Conference}} on
  {{Software Engineering}}}, ICSE '07, pages 261--270, Washington, DC, USA,
  2007. {IEEE Computer Society}.

\bibitem{Czarnecki:ICSM'12}
K.~Czarnecki, Z.~Malik, and R.~Lotufo.
\newblock Modelling the \&\#8216;{{Hurried}}\&\#8217; {{Bug Report Reading
  Process}} to {{Summarize Bug Reports}}.
\newblock In {\em Proceedings of the 2012 {{IEEE International Conference}} on
  {{Software Maintenance}} ({{ICSM}})}, ICSM '12, pages 430--439, Washington,
  DC, USA, 2012. {IEEE Computer Society}.

\bibitem{Deng:ICSTW'15}
L.~Deng, N.~Mirzaei, P.~Ammann, and J.~Offutt.
\newblock Towards mutation analysis of {{Android}} apps.
\newblock In {\em {{ICSTW}} '15}, ICSTW '15, pages 1--10, Apr. 2015.

\bibitem{Deng:IST'17}
L.~Deng, J.~Offutt, P.~Ammann, and N.~Mirzaei.
\newblock Mutation {{Operators}} for {{Testing Android Apps}}.
\newblock {\em Inf. Softw. Technol.}, 81(C):154--168, Jan. 2017.

\bibitem{DiNucci:ICSE'17C}
D.~Di~Nucci, F.~Palomba, A.~Prota, A.~Panichella, A.~Zaidman, and A.~De~Lucia.
\newblock {PET}r{A}: A software-based tool for estimating the energy profile of
  android applications.
\newblock In {\em Proceedings of the 39th International Conference on Software
  Engineering Companion}, ICSE-C'17, pages 3--6, Piscataway, NJ, USA, 2017.
  IEEE Press.

\bibitem{Dit:EMSE'13}
B.~Dit, M.~Revelle, M.~Gethers, and D.~Poshyvanyk.
\newblock Feature location in source code: a taxonomy and survey.
\newblock {\em Journal of Software: Evolution and Process}, 25(1):53--95.

\bibitem{Dong:TMC'12-2}
M.~Dong and L.~Zhong.
\newblock Chameleon: A color-adaptive web browser for mobile {OLED} displays.
\newblock {\em IEEE Transactions on Mobile Computing}, 11(5):724--738, May
  2012.

\bibitem{Dong:TMC'12}
M.~Dong and L.~Zhong.
\newblock Power modeling and optimization for {OLED} displays.
\newblock {\em IEEE Transaction on Mobile Computing}, 11(9):September, 2012.

\bibitem{Fazzini:ICST'17}
M.~Fazzini, E.~N. D.~A. Freitas, S.~R. Choudhary, and A.~Orso.
\newblock Barista: {{A Technique}} for {{Recording}}, {{Encoding}}, and
  {{Running Platform Independent Android Tests}}.
\newblock In {\em 2017 {{IEEE International Conference}} on {{Software
  Testing}}, {{Verification}} and {{Validation}} ({{ICST}})}, ICST'17, pages
  149--160, Mar. 2017.
\newblock ISSN:.

\bibitem{Fazzini:ASE'17}
M.~Fazzini and A.~Orso.
\newblock Automated cross-platform inconsistency detection for mobile apps.
\newblock In {\em 2017 32nd {{IEEE}}/{{ACM International Conference}} on
  {{Automated Software Engineering}} ({{ASE}})}, ASE'17, pages 308--318, Oct.
  2017.
\newblock ISSN:.

\bibitem{Gomez:ICSE'13}
L.~Gomez, I.~Neamtiu, T.~Azim, and T.~Millstein.
\newblock {{RERAN}}: {{Timing}}- and {{Touch}}-sensitive {{Record}} and
  {{Replay}} for {{Android}}.
\newblock In {\em Proceedings of the 2013 {{International Conference}} on
  {{Software Engineering}}}, ICSE'13, pages 72--81, San Francisco, CA, USA,
  2013. {IEEE Press}.

\bibitem{Grano:SANER'18}
G.~Grano, A.~Ciurumelea, S.~Panichella, S.~Palomba, and H.~Gall.
\newblock Exploring the integration of user feedback in automated testing of
  android applications.
\newblock In {\em Proceedings of the 25th IEEE International Conference on
  Software Analysis, Evolution and Reengineering}, SANER '18, 2018.

\bibitem{Gu:ICSME'17}
T.~Gu, C.~Cao, T.~Liu, C.~Sun, J.~Deng, X.~Ma, and J.~L{\"u}.
\newblock {{AimDroid}}: {{Activity}}-{{Insulated Multi}}-level {{Automated
  Testing}} for {{Android Applications}}.
\newblock In {\em 2017 {{IEEE International Conference}} on {{Software
  Maintenance}} and {{Evolution}} ({{ICSME}})}, ICSME'17, pages 103--114, Sept.
  2017.
\newblock ISSN:.

\bibitem{Gu:ICSE'10}
Z.~Gu, E.~Barr, D.~Hamilton, and Z.~Su.
\newblock Has the bug really been fixed?
\newblock In {\em Software {{Engineering}}, 2010 {{ACM}}/{{IEEE}} 32nd
  {{International Conference}} On}, volume~1 of {\em ICSE'10}, pages 55--64,
  May 2010.

\bibitem{Gui:ICSE'15}
J.~Gui, S.~Mcilroy, M.~Nagappan, and W.~G.~J. Halfond.
\newblock Truth in {{Advertising}}: {{The Hidden Cost}} of {{Mobile Ads}} for
  {{Software Developers}}.
\newblock In {\em Proceedings of the 37th {{International Conference}} on
  {{Software Engineering}} - {{Volume}} 1}, ICSE '15, pages 100--110, Florence,
  Italy, 2015. {IEEE Press}.

\bibitem{Guo:ICSE'10}
P.~J. Guo, T.~Zimmermann, N.~Nagappan, and B.~Murphy.
\newblock Characterizing and {{Predicting Which Bugs Get Fixed}}: {{An
  Empirical Study}} of {{Microsoft Windows}}.
\newblock In {\em Proceedings of the {{32Nd ACM}}/{{IEEE International
  Conference}} on {{Software Engineering}} - {{Volume}} 1}, ICSE '10, pages
  495--504, Cape Town, South Africa, 2010. {ACM}.

\bibitem{Han:WCRE'12}
D.~Han, C.~Zhang, X.~Fan, A.~Hindle, K.~Wong, and E.~Stroulia.
\newblock Understanding android fragmentation with topic analysis of
  vendor-specific bugs.
\newblock In {\em Proceedings of the 2012 19th Working Conference on Reverse
  Engineering}, WCRE '12, pages 83--92, Washington, DC, USA, 2012. IEEE
  Computer Society.

\bibitem{Hao:MobiSys'14}
S.~Hao, B.~Liu, S.~Nath, W.~G. Halfond, and R.~Govindan.
\newblock {{PUMA}}: {{Programmable UI}}-automation for {{Large}}-scale
  {{Dynamic Analysis}} of {{Mobile Apps}}.
\newblock In {\em Proceedings of the 12th {{Annual International Conference}}
  on {{Mobile Systems}}, {{Applications}}, and {{Services}}}, MobiSys '14,
  pages 204--217, Bretton Woods, New Hampshire, USA, 2014. {ACM}.

\bibitem{Hu:EuroSys'14}
G.~Hu, X.~Yuan, Y.~Tang, and J.~Yang.
\newblock Efficiently, {{Effectively Detecting Mobile App Bugs}} with
  {{AppDoctor}}.
\newblock In {\em Proceedings of the {{Ninth European Conference}} on
  {{Computer Systems}}}, EuroSys '14, pages 18:1--18:15, Amsterdam, The
  Netherlands, 2014. {ACM}.

\bibitem{Hu:OOPSLA2015}
Y.~Hu, T.~Azim, and I.~Neamtiu.
\newblock Versatile {{Yet Lightweight Record}}-and-replay for {{Android}}.
\newblock In {\em {{OOPSLA}}'15}, OOPSLA 2015, pages 349--366, Pittsburgh, PA,
  USA, 2015. {ACM}.

\bibitem{Jabbarvand:GREENS'15}
R.~Jabbarvand, A.~Sadeghi, J.~Garcia, S.~Malek, and P.~Ammann.
\newblock Ecodroid: An approach for energy-based ranking of android apps.
\newblock In {\em Proceedings of the Fourth International Workshop on Green and
  Sustainable Software}, GREENS '15, pages 8--14, Piscataway, NJ, USA, 2015.
  IEEE Press.

\bibitem{Jeong:ESEC/FSE'09}
G.~Jeong, S.~Kim, and T.~Zimmermann.
\newblock Improving {{Bug Triage}} with {{Bug Tossing Graphs}}.
\newblock In {\em Proceedings of the the 7th {{Joint Meeting}} of the
  {{European Software Engineering Conference}} and the {{ACM SIGSOFT
  Symposium}} on {{The Foundations}} of {{Software Engineering}}}, ESEC/FSE
  '09, pages 111--120, Amsterdam, The Netherlands, 2009. {ACM}.

\bibitem{Jin:ICSE'12}
W.~Jin and A.~Orso.
\newblock {{BugRedux}}: {{Reproducing Field Failures}} for {{In}}-house
  {{Debugging}}.
\newblock In {\em Proceedings of the 34th {{International Conference}} on
  {{Software Engineering}}}, ICSE '12, pages 474--484, Zurich, Switzerland,
  2012. {IEEE Press}.

\bibitem{Jin:ISSTA'13}
W.~Jin and A.~Orso.
\newblock F3: {{Fault Localization}} for {{Field Failures}}.
\newblock In {\em Proceedings of the 2013 {{International Symposium}} on
  {{Software Testing}} and {{Analysis}}}, ISSTA'13, pages 213--223, Lugano,
  Switzerland, 2013. {ACM}.

\bibitem{Jones:2014}
N.~Jones.
\newblock Seven best practices for optimizing mobile testing efforts.
\newblock Technical Report G00248240, {Gartner}.

\bibitem{Joorabchi:MSR'14}
M.~Joorabchi, M.~Mirzaaghaei, and A.~Mesbah.
\newblock Works for {{Me}}! {{Characterizing Non}}-reproducible {{Bug
  Reports}}.
\newblock In {\em Proceedings of the 11th {{Working Conference}} on {{Mining
  Software Repositories}}}, MSR'14, pages 62--71, Hyderabad, India, 2014.
  {ACM}.

\bibitem{Joorabchi:ISSRE'15}
M.~E. Joorabchi, M.~Ali, and A.~Mesbah.
\newblock Detecting inconsistencies in multi-platform mobile apps.
\newblock In {\em 2015 {{IEEE}} 26th {{International Symposium}} on {{Software
  Reliability Engineering}} ({{ISSRE}})}, ISSRE'15, pages 450--460, Nov. 2015.

\bibitem{Khalid:FSE'14}
H.~Khalid, M.~Nagappan, E.~Shihab, and A.~E. Hassan.
\newblock Prioritizing the {{Devices}} to {{Test Your App}} on: {{A Case
  Study}} of {{Android Game Apps}}.
\newblock In {\em Proceedings of the {{22Nd ACM SIGSOFT International
  Symposium}} on {{Foundations}} of {{Software Engineering}}}, FSE'14, pages
  610--620, Hong Kong, China, 2014. {ACM}.

\bibitem{Kifetew:ICST'14}
F.~M. Kifetew, W.~Jin, R.~Tiella, A.~Orso, and P.~Tonella.
\newblock Reproducing {{Field Failures}} for {{Programs}} with {{Complex
  Grammar}}-{{Based Input}}.
\newblock In {\em 2014 {{IEEE Seventh International Conference}} on {{Software
  Testing}}, {{Verification}} and {{Validation}}}, ICST'14, pages 163--172,
  Mar. 2014.

\bibitem{44Kim:TOSE2013}
D.~Kim, Y.~Tao, S.~Kim, and A.~Zeller.
\newblock Where {{Should We Fix This Bug}}? {{A Two}}-{{Phase Recommendation
  Model}}.
\newblock {\em Software Engineering, IEEE Transactions on}, 39(11):1597--1610,
  Nov. 2013.

\bibitem{Kim:DSN'11}
S.~Kim, T.~Zimmermann, and N.~Nagappan.
\newblock Crash graphs: {{An}} aggregated view of multiple crashes to improve
  crash triage.
\newblock In {\em Dependable {{Systems Networks}} ({{DSN}}), 2011
  {{IEEE}}/{{IFIP}} 41st {{International Conference}} On}, DSN'11, pages
  486--493, June 2011.

\bibitem{33Koru:IEEE2004}
A.~G. Koru and J.~Tian.
\newblock Defect {{Handling}} in {{Medium}} and {{Large Open Source Projects}}.
\newblock {\em IEEE Softw.}, 21(4):54--61, July 2004.

\bibitem{Lelli:ICST'15}
V.~Lelli, A.~Blouin, and B.~Baudry.
\newblock Classifying and {{Qualifying GUI Defects}}.
\newblock In {\em 2015 {{IEEE}} 8th {{International Conference}} on {{Software
  Testing}}, {{Verification}} and {{Validation}} ({{ICST}})}, ICST'15, pages
  1--10, Apr. 2015.

\bibitem{Li:DeMobile'15}
D.~Li and W.~G.~J. Halfond.
\newblock Optimizing energy of http requests in android applications.
\newblock In {\em Proceedings of the 3rd International Workshop on Software
  Development Lifecycle for Mobile}, DeMobile 2015, pages 25--28, 2015.

\bibitem{Li:ICSME'14}
D.~Li, S.~Hao, J.~Gui, and W.~G.~J. Halfond.
\newblock An {{Empirical Study}} of the {{Energy Consumption}} of {{Android
  Applications}}.
\newblock In {\em 2014 {{IEEE International Conference}} on {{Software
  Maintenance}} and {{Evolution}}}, ICSME'14, pages 121--130, Sept. 2014.

\bibitem{Li:ISSTA'13}
D.~Li, S.~Hao, W.~G.~J. Halfond, and R.~Govindan.
\newblock Calculating {{Source Line Level Energy Information}} for {{Android
  Applications}}.
\newblock In {\em Proceedings of the 2013 {{International Symposium}} on
  {{Software Testing}} and {{Analysis}}}, ISSTA'13, pages 78--89, Lugano,
  Switzerland, 2013. {ACM}.

\bibitem{Li:ICSE'16}
D.~Li, Y.~Lyu, J.~Gui, and W.~G.~J. Halfond.
\newblock Automated {{Energy Optimization}} of {{HTTP Requests}} for {{Mobile
  Applications}}.
\newblock In {\em Proceedings of the 38th {{International Conference}} on
  {{Software Engineering}}}, ICSE '16, pages 249--260, New York, NY, USA, 2016.
  {ACM}.

\bibitem{Li:ICSE'14}
D.~Li, A.~H. Tran, and W.~G.~J. Halfond.
\newblock Making {{Web Applications More Energy Efficient}} for {{OLED
  Smartphones}}.
\newblock In {\em Proceedings of the 36th {{International Conference}} on
  {{Software Engineering}}}, ICSE'14, pages 527--538, Hyderabad, India, 2014.
  {ACM}.

\bibitem{Li:IST'17}
L.~Li, T.~F. Bissyandé, M.~Papadakis, S.~Rasthofer, A.~Bartel, D.~Octeau,
  J.~Klein, and L.~Traon.
\newblock Static analysis of android apps: A systematic literature review.
\newblock {\em Information and Software Technology}, 88:67 -- 95, 2017.

\bibitem{Lin:ASE'15b}
Y.~Lin, S.~Okur, and D.~Dig.
\newblock Study and {{Refactoring}} of {{Android Asynchronous Programming}}
  ({{T}}).
\newblock In {\em 2015 30th {{IEEE}}/{{ACM International Conference}} on
  {{Automated Software Engineering}} ({{ASE}})}, ASE'15, pages 224--235, Nov.
  2015.
\newblock ISSN:.

\bibitem{Linares-Vasquez:FSE'13}
M.~Linares-V{\'a}squez, G.~Bavota, C.~Bernal-C{\'a}rdenas, M.~Di~Penta,
  R.~Oliveto, and D.~Poshyvanyk.
\newblock {{API Change}} and {{Fault Proneness}}: {{A Threat}} to the
  {{Success}} of {{Android Apps}}.
\newblock In {\em Proceedings of the 2013 9th {{Joint Meeting}} on
  {{Foundations}} of {{Software Engineering}}}, FSE'13, pages 477--487, Saint
  Petersburg, Russia, 2013. {ACM}.

\bibitem{Linares-Vasquez:MSR'14}
M.~Linares-V{\'a}squez, G.~Bavota, C.~Bernal-C{\'a}rdenas, R.~Oliveto,
  M.~Di~Penta, and D.~Poshyvanyk.
\newblock Mining {{Energy}}-greedy {{API Usage Patterns}} in {{Android Apps}}:
  {{An Empirical Study}}.
\newblock In {\em Proceedings of the 11th {{Working Conference}} on {{Mining
  Software Repositories}}}, MSR'14, pages 2--11, Hyderabad, India, 2014. {ACM}.

\bibitem{Linares-Vasquez:FSE'15}
M.~Linares-V{\'a}squez, G.~Bavota, C.~E.~B. C{\'a}rdenas, R.~Oliveto,
  M.~Di~Penta, and D.~Poshyvanyk.
\newblock Optimizing {{Energy Consumption}} of {{GUIs}} in {{Android Apps}}:
  {{A Multi}}-objective {{Approach}}.
\newblock In {\em Proceedings of the 2015 10th {{Joint Meeting}} on
  {{Foundations}} of {{Software Engineering}}}, FSE'15, pages 143--154,
  Bergamo, Italy, 2015. {ACM}.

\bibitem{Linares-Vasquez:FSE'17}
M.~Linares-V{\'a}squez, G.~Bavota, M.~Tufano, K.~Moran, M.~Di~Penta,
  C.~Vendome, C.~Bernal-C{\'a}rdenas, and D.~Poshyvanyk.
\newblock Enabling {{Mutation Testing}} for {{Android Apps}}.
\newblock In {\em Proceedings of the 2017 11th {{Joint Meeting}} on
  {{Foundations}} of {{Software Engineering}}}, FSE'17, pages 233--244,
  Paderborn, Germany, 2017. {ACM}.

\bibitem{Linares-Vasquez:ICSME'17a}
M.~Linares-V{\'a}squez, C.~Bernal-Cardenas, K.~Moran, and D.~Poshyvanyk.
\newblock How do {{Developers Test Android Applications}}?
\newblock In {\em 2017 {{IEEE International Conference}} on {{Software
  Maintenance}} and {{Evolution}} ({{ICSME}})}, ICSME'17, pages 613--622, Sept.
  2017.
\newblock ISSN:.

\bibitem{Linares-Vasquez:a}
M.~Linares-V{\'a}squez, K.~Hossen, H.~Dang, H.~Kagdi, M.~Gethers, and
  D.~Poshyvanyk.
\newblock Triaging incoming change requests: {{Bug}} or commit history, or code
  authorship?
\newblock In {\em Software {{Maintenance}} ({{ICSM}}), 2012 28th {{IEEE
  International Conference}} On}, pages 451--460, Sept. 2012.

\bibitem{Linares-Vasquez:ICSME'17}
M.~Linares-V{\'a}squez, K.~Moran, and D.~Poshyvanyk.
\newblock Continuous, {{Evolutionary}} and {{Large}}-{{Scale}}: {{A New
  Perspective}} for {{Automated Mobile App Testing}}.
\newblock In {\em 2017 {{IEEE International Conference}} on {{Software
  Maintenance}} and {{Evolution}} ({{ICSME}})}, ICSME'17, pages 399--410, Sept.
  2017.
\newblock ISSN:.

\bibitem{Linares-Vasquez:ICSME'15}
M.~Linares-V{\'a}squez, C.~Vendome, Q.~Luo, and D.~Poshyvanyk.
\newblock How developers detect and fix performance bottlenecks in {{Android}}
  apps.
\newblock In {\em 2015 {{IEEE International Conference}} on {{Software
  Maintenance}} and {{Evolution}} ({{ICSME}})}, ICSME'15, pages 352--361, Sept.
  2015.
\newblock ISSN:.

\bibitem{Linares-Vasquez:MSR'15}
M.~Linares-V{\'a}squez, M.~White, C.~Bernal-C{\'a}rdenas, K.~Moran, and
  D.~Poshyvanyk.
\newblock Mining {{Android App Usages}} for {{Generating Actionable GUI}}-based
  {{Execution Scenarios}}.
\newblock In {\em Proceedings of the 12th {{Working Conference}} on {{Mining
  Software Repositories}}}, MSR '15, pages 111--122, Florence, Italy, 2015.
  {IEEE Press}.

\bibitem{Linares-Vasquez:ICSE-C'17}
M.~Linares-Vásquez, C.~Bernal-Cárdenas, G.~Bavota, R.~Oliveto, M.~D. Penta,
  and D.~Poshyvanyk.
\newblock Gemma: Multi-objective optimization of energy consumption of guis in
  android apps.
\newblock In {\em 2017 IEEE/ACM 39th International Conference on Software
  Engineering Companion (ICSE-C)}, ICSE-C'17, 2017.

\bibitem{Liu:TSE'14}
Y.~Liu, C.~Xu, S.~Cheung, and J.~Lu.
\newblock Green{D}roid: Automated diagnosis of energy inefficiency for
  smartphone applications.
\newblock {\em IEEE Transactions on Software Engineering}, Preprint, 2014.

\bibitem{Lu:ICSE'16a}
X.~Lu, X.~Liu, H.~Li, T.~Xie, Q.~Mei, D.~Hao, G.~Huang, and F.~Feng.
\newblock {{PRADA}}: {{Prioritizing Android Devices}} for {{Apps}} by {{Mining
  Large}}-scale {{Usage Data}}.
\newblock In {\em Proceedings of the 38th {{International Conference}} on
  {{Software Engineering}}}, ICSE '16, pages 3--13, New York, NY, USA, 2016.
  {ACM}.

\bibitem{Machiry:FSE'13}
A.~Machiry, R.~Tahiliani, and M.~Naik.
\newblock Dynodroid: {{An Input Generation System}} for {{Android Apps}}.
\newblock In {\em Proceedings of the 2013 9th {{Joint Meeting}} on
  {{Foundations}} of {{Software Engineering}}}, FSE'13, pages 224--234, Saint
  Petersburg, Russia, 2013. {ACM}.

\bibitem{Mahajan:ICSE'18}
S.~Mahajan, N.~Abolhasani, P.~McMinn, and W.~G. Halfond.
\newblock Automated repair of mobile friendly problems in web pages.
\newblock In {\em Proceedings of the International Conference on Software
  Engineering (ICSE)}, May 2018.
\newblock To Appear.

\bibitem{Mani:FSE'12}
S.~Mani, R.~Catherine, V.~S. Sinha, and A.~Dubey.
\newblock {{AUSUM}}: {{Approach}} for {{Unsupervised Bug Report
  Summarization}}.
\newblock In {\em Proceedings of the {{ACM SIGSOFT}} 20th {{International
  Symposium}} on the {{Foundations}} of {{Software Engineering}}}, FSE '12,
  pages 11:1--11:11, Cary, North Carolina, 2012. {ACM}.

\bibitem{Mao:ISSTA'16}
K.~Mao, M.~Harman, and Y.~Jia.
\newblock Sapienz: {{Multi}}-objective {{Automated Testing}} for {{Android
  Applications}}.
\newblock In {\em Proceedings of the 25th {{International Symposium}} on
  {{Software Testing}} and {{Analysis}}}, ISSTA'16, pages 94--105,
  Saarbr\&\#252;cken, Germany, 2016. {ACM}.

\bibitem{Mao:ASE'17}
K.~Mao, M.~Harman, and Y.~Jia.
\newblock Crowd intelligence enhances automated mobile testing.
\newblock In {\em 2017 32nd {{IEEE}}/{{ACM International Conference}} on
  {{Automated Software Engineering}} ({{ASE}})}, ASE'17, pages 16--26, Oct.
  2017.
\newblock ISSN:.

\bibitem{Martin:TSE'17}
W.~Martin, F.~Sarro, Y.~Jia, Y.~Zhang, and M.~Harman.
\newblock A survey of app store analysis for software engineering.
\newblock {\em IEEE transactions on software engineering}, 43(9):817--847,
  2017.

\bibitem{McDonnell:ICSM'13}
T.~McDonnell, B.~Ray, and M.~Kim.
\newblock An {{Empirical Study}} of {{API Stability}} and {{Adoption}} in the
  {{Android Ecosystem}}.
\newblock In {\em Proceedings of the 2013 {{International Conference}} on
  {{Software Maintenance}}}, ICSM'13, pages 70--79, 2013.

\bibitem{Moran:ArX'18}
K.~{Moran}, C.~{Bernal-C{\'a}rdenas}, M.~{Curcio}, R.~{Bonett}, and
  D.~{Poshyvanyk}.
\newblock {Machine Learning-Based Prototyping of Graphical User Interfaces for
  Mobile Apps}.
\newblock {\em ArXiv e-prints}, Feb. 2018.

\bibitem{Moran:MobileSoft'17}
K.~Moran, R.~Bonett, C.~Bernal-C{\'a}rdenas, B.~Otten, D.~Park, and
  D.~Poshyvanyk.
\newblock On-{{Device Bug Reporting}} for {{Android Applications}}.
\newblock In {\em {{MobileSOFT}}'17}, MobileSoft'17, May 2017.

\bibitem{Moran:ICSE'18}
K.~Moran, B.~Li, C.~Bernal-C{\'a}rdenas, D.~Jelf, and D.~Poshyvanyk.
\newblock Automated {{Reporting}} of {{GUI Design Violations}} in {{Mobile
  Apps}}.
\newblock In {\em Proceedings of the 40th {{International Conference}} on
  {{Software Engineering Companion}}}, ICSE '18, page to appear, Gothenburg,
  Sweden, 2018. {IEEE Press}.

\bibitem{Moran:FSE'15}
K.~Moran, M.~Linares-V{\'a}squez, C.~Bernal-C{\'a}rdenas, and D.~Poshyvanyk.
\newblock Auto-completing {{Bug Reports}} for {{Android Applications}}.
\newblock In {\em Proceedings of the 2015 10th {{Joint Meeting}} on
  {{Foundations}} of {{Software Engineering}}}, FSE'15, pages 673--686,
  Bergamo, Italy, 2015. {ACM}.

\bibitem{Moran:ICSE'16}
K.~Moran, M.~Linares-V{\'a}squez, C.~Bernal-C{\'a}rdenas, and D.~Poshyvanyk.
\newblock {{FUSION}}: {{A Tool}} for {{Facilitating}} and {{Augmenting Android
  Bug Reporting}}.
\newblock In {\em {{ICSE}}'16}, ICSE'16, May 2016.

\bibitem{Moran:ICST'16}
K.~Moran, M.~Linares-V{\'a}squez, C.~Bernal-C{\'a}rdenas, C.~Vendome, and
  D.~Poshyvanyk.
\newblock Automatically {{Discovering}}, {{Reporting}} and {{Reproducing
  Android Application Crashes}}.
\newblock In {\em 2016 {{IEEE International Conference}} on {{Software
  Testing}}, {{Verification}} and {{Validation}} ({{ICST}})}, ICST'16, pages
  33--44, Apr. 2016.
\newblock ISSN:.

\bibitem{Moran:ICSE'18-2}
K.~Moran, M.~Tufano, C.~Bernal-C{\'a}rdenas, M.~Linares-V{\'a}squez, G.~Bavota,
  C.~Vendome, M.~Di~Penta, and D.~Poshyvanyk.
\newblock Mdroid+: A mutation testing framework for android.
\newblock In {\em Proceedings of the 40th {{International Conference}} on
  {{Software Engineering Companion}}}, ICSE '18, page to appear, Gothenburg,
  Sweden, 2018. {IEEE Press}.

\bibitem{Muccini:AST'12}
H.~Muccini, A.~Di~Francesco, and P.~Esposito.
\newblock Software testing of mobile applications: Challenges and future
  research directions.
\newblock In {\em Proceedings of the 7th International Workshop on Automation
  of Software Test}, AST '12, pages 29--35, Piscataway, NJ, USA, 2012. IEEE
  Press.

\bibitem{Myers:CHD94}
B.~Myers.
\newblock Challenges of {{HCI Design}} and {{Implementation}}.
\newblock {\em Interactions}, 1(1):73--83, Jan. 1994.

\bibitem{Naguib:MSR'13}
H.~Naguib, N.~Narayan, B.~Br{\"u}gge, and D.~Helal.
\newblock Bug {{Report Assignee Recommendation Using Activity Profiles}}.
\newblock In {\em Proceedings of the 10th {{Working Conference}} on {{Mining
  Software Repositories}}}, MSR '13, pages 22--30, San Francisco, CA, USA,
  2013. {IEEE Press}.

\bibitem{Nguyen:ASE'15}
A.~T. Nguyen, T.~T. Nguyen, and T.~N. Nguyen.
\newblock Divide-and-{{Conquer Approach}} for {{Multi}}-phase {{Statistical
  Migration}} for {{Source Code}} ({{T}}).
\newblock In {\em 2015 30th {{IEEE}}/{{ACM International Conference}} on
  {{Automated Software Engineering}} ({{ASE}})}, ASE'15, pages 585--596, Nov.
  2015.
\newblock ISSN:.

\bibitem{Nguyen:ASE2012}
A.~T. Nguyen, T.~T. Nguyen, T.~N. Nguyen, D.~Lo, and C.~Sun.
\newblock Duplicate {{Bug Report Detection}} with a {{Combination}} of
  {{Information Retrieval}} and {{Topic Modeling}}.
\newblock In {\em Proceedings of the 27th {{IEEE}}/{{ACM International
  Conference}} on {{Automated Software Engineering}}}, ASE 2012, pages 70--79,
  Essen, Germany, 2012. {ACM}.

\bibitem{Nguyen:ASE'15b}
T.~A. Nguyen and C.~Csallner.
\newblock Reverse {{Engineering Mobile Application User Interfaces}} with
  {{REMAUI}}.
\newblock In {\em Proceedings of the 2015 30th {{IEEE}}/{{ACM International
  Conference}} on {{Automated Software Engineering}}}, ASE'15, pages 248--259,
  Washington, DC, USA, 2015. {IEEE Computer Society}.

\bibitem{DiNucci:SANER'17}
D.~D. Nucci, F.~Palomba, A.~Prota, A.~Panichella, A.~Zaidman, and A.~D. Lucia.
\newblock Software-based energy profiling of android apps: Simple, efficient
  and reliable?
\newblock In {\em 2017 IEEE 24th International Conference on Software Analysis,
  Evolution and Reengineering}, SANER'17, pages 103--114, Feb 2017.

\bibitem{Palomba:ICSME'15}
F.~Palomba, M.~Linares-V{\'a}squez, G.~Bavota, R.~Oliveto, M.~D. Penta,
  D.~Poshyvanyk, and A.~D. Lucia.
\newblock User reviews matter! {{Tracking}} crowdsourced reviews to support
  evolution of successful apps.
\newblock In {\em 2015 {{IEEE International Conference}} on {{Software
  Maintenance}} and {{Evolution}} ({{ICSME}})}, ICSME'15, pages 291--300, Sept.
  2015.
\newblock ISSN:.

\bibitem{Palomba:JSS17}
F.~Palomba, M.~Linares-Vásquez, G.~Bavota, R.~Oliveto, M.~D. Penta,
  D.~Poshyvanyk, and A.~D. Lucia.
\newblock Crowdsourcing user reviews to support the evolution of mobile apps.
\newblock {\em Journal of Systems and Software}, 137:143 -- 162, 2018.

\bibitem{Palomba:ICSE'17}
F.~Palomba, P.~Salza, A.~Ciurumelea, S.~Panichella, H.~Gall, F.~Ferrucci, and
  A.~D. Lucia.
\newblock Recommending and {{Localizing Code Changes}} for {{Mobile Apps}}
  based on {{User Reviews}}.
\newblock In {\em {{ICSE}}'17}, 2017.

\bibitem{Pathak:Hotnets'11}
A.~Pathak, Y.~Hu, and M.~Zhang.
\newblock Bootstrapping {{Energy Debugging}} on {{Smartphones}}: {{A First
  Look}} at {{Energy Bugs}} in {{Mobile Devices}}.
\newblock In {\em Hotnets'11}, Hotnets'11.

\bibitem{Pathak:EuroSys'12}
A.~Pathak, Y.~Hu, and M.~Zhang.
\newblock Where is the energy spent inside my app? {{Fine Grained Energy
  Accounting}} on {{Smartphones}} with {{Eprof}}.
\newblock In {\em {{EuroSys}}'12}, EuroSys'12, pages 29--42, 2012.

\bibitem{Pathak:MobiSys'12}
A.~Pathak, A.~Jindal, Y.~Hu, and S.~P. Midkiff.
\newblock What is keeping my phone awake? {{Characterizing}} and {{Detecting
  No}}-{{Sleep Energy Bugs}} in {{Smartphone Apps}}.
\newblock In {\em {{MobiSys}}'12}, MobiSys'12, pages 267--280, 2012.

\bibitem{Rasmussen:GREENS'14}
K.~Rasmussen, A.~Wilson, and A.~Hindle.
\newblock Green mining: energy consumption of advertisement blocking methods.
\newblock In {\em GREENS'14}, pages 38--45, 2014.

\bibitem{Rastkar:ICSE'10}
S.~Rastkar, G.~C. Murphy, and G.~Murray.
\newblock Summarizing {{Software Artifacts}}: {{A Case Study}} of {{Bug
  Reports}}.
\newblock In {\em Proceedings of the {{32Nd ACM}}/{{IEEE International
  Conference}} on {{Software Engineering}} - {{Volume}} 1}, ICSE '10, pages
  505--514, Cape Town, South Africa, 2010. {ACM}.

\bibitem{Robillard:ICSME'17}
M.~P. Robillard, A.~Marcus, C.~Treude, G.~Bavota, O.~Chaparro, N.~Ernst, M.~A.
  Gerosa, M.~Godfrey, M.~Lanza, M.~Linares-V{\'a}squez, G.~C. Murphy,
  L.~Moreno, D.~Shepherd, and E.~Wong.
\newblock On-demand {{Developer Documentation}}.
\newblock In {\em 2017 {{IEEE International Conference}} on {{Software
  Maintenance}} and {{Evolution}} ({{ICSME}})}, ICSME'17, pages 479--483, Sept.
  2017.
\newblock ISSN:.

\bibitem{Romansky:ICSME'17}
S.~Romansky, N.~C. Borle, S.~Chowdhury, A.~Hindle, and R.~Greiner.
\newblock Deep {{Green}}: {{Modelling Time}}-{{Series}} of {{Software Energy
  Consumption}}.
\newblock In {\em 2017 {{IEEE International Conference}} on {{Software
  Maintenance}} and {{Evolution}} ({{ICSME}})}, ICSME'17, pages 273--283, Sept.
  2017.
\newblock ISSN:.

\bibitem{Saborino:CoRR'17}
R.~Saborido, F.~Khomh, A.~Hindle, and E.~Alba.
\newblock An app performance optimization advisor for mobile device app
  marketplaces.
\newblock {\em CoRR}, abs/1709.04916, 2017.

\bibitem{Sadeghi:TSE'17}
A.~Sadeghi, H.~Bagheri, J.~Garcia, and S.~Malek.
\newblock A taxonomy and qualitative comparison of program analysis techniques
  for security assessment of android software.
\newblock {\em IEEE Transactions on Software Engineering}, 43(6):492--530, June
  2017.

\bibitem{Sahin:JSEP'16}
C.~Sahin, M.~Wan, P.~Tornquist, R.~McKenna, Z.~Pearson, W.~G.~J. Halfond, and
  J.~Clause.
\newblock How does code obfuscation impact energy usage?
\newblock {\em Journal of Software: Evolution and Process}, pages n/a--n/a,
  2016.

\bibitem{Sasnauskas:WODA14}
R.~Sasnauskas and J.~Regehr.
\newblock Intent {{Fuzzer}}: {{Crafting Intents}} of {{Death}}.
\newblock In {\em Proceedings of the 2014 {{Joint International Workshop}} on
  {{Dynamic Analysis}} and {{Software}} and {{System Performance Testing}},
  {{Debugging}}, and {{Analytics}}}, WODA+PERTEA'14, pages 1--5, San Jose, CA,
  USA, 2014. {ACM}.

\bibitem{Shokripour:MSR'13}
R.~Shokripour, J.~Anvik, Z.~M. Kasirun, and S.~Zamani.
\newblock Why {{So Complicated}}? {{Simple Term Filtering}} and {{Weighting}}
  for {{Location}}-based {{Bug Report Assignment Recommendation}}.
\newblock In {\em Proceedings of the 10th {{Working Conference}} on {{Mining
  Software Repositories}}}, MSR '13, pages 2--11, San Francisco, CA, USA, 2013.
  {IEEE Press}.

\bibitem{Stanley:CC'16}
P.~Stanley-Marbell, V.~Estellers, and M.~Rinard.
\newblock Crayon: saving power through shape and color approximation on
  next-generation displays.
\newblock In {\em Proceedings of the Eleventh European Conference on Computer
  Systems}, page~11. ACM, 2016.

\bibitem{Tucker:CSH04}
A.~B. Tucker.
\newblock {\em Computer {{Science Handbook}}, {{Second Edition}}}.
\newblock {Chapman \& Hall/CRC}, 2004.

\bibitem{Wan:ICST'15}
M.~Wan, Y.~Jin, D.~Li, and W.~G.~J. Halfond.
\newblock Detecting {{Display Energy Hotspots}} in {{Android Apps}}.
\newblock In {\em 2015 {{IEEE}} 8th {{International Conference}} on {{Software
  Testing}}, {{Verification}} and {{Validation}} ({{ICST}})}, ICST'15, pages
  1--10, Apr. 2015.

\bibitem{Wang:ICSE'08}
X.~Wang, L.~Zhang, T.~Xie, J.~Anvik, and J.~Sun.
\newblock An {{Approach}} to {{Detecting Duplicate Bug Reports Using Natural
  Language}} and {{Execution Information}}.
\newblock In {\em Proceedings of the 30th {{International Conference}} on
  {{Software Engineering}}}, ICSE '08, pages 461--470, Leipzig, Germany, 2008.
  {ACM}.

\bibitem{Wei:ASE'16}
L.~Wei, Y.~Liu, and S.~C. Cheung.
\newblock Taming {{Android}} fragmentation: {{Characterizing}} and detecting
  compatibility issues for {{Android}} apps.
\newblock In {\em 2016 31st {{IEEE}}/{{ACM International Conference}} on
  {{Automated Software Engineering}} ({{ASE}})}, ASE'16, pages 226--237, Sept.
  2016.
\newblock ISSN:.

\bibitem{Weiss:MSR'07}
C.~Weiss, R.~Premraj, T.~Zimmermann, and A.~Zeller.
\newblock How {{Long Will It Take}} to {{Fix This Bug}}?
\newblock In {\em Proceedings of the {{Fourth International Workshop}} on
  {{Mining Software Repositories}}}, MSR '07, pages 1--, Washington, DC, USA,
  2007. {IEEE Computer Society}.

\bibitem{Wu:CC'16}
H.~Wu, S.~Yang, and A.~Rountev.
\newblock Static detection of energy defect patterns in android applications.
\newblock In {\em Proceedings of the 25th International Conference on Compiler
  Construction}, CC'16, pages 185--195, New York, NY, USA, 2016. ACM.

\bibitem{Yang:FASE'13}
W.~Yang, M.~R. Prasad, and T.~Xie.
\newblock A {{Grey}}-box {{Approach}} for {{Automated GUI}}-model
  {{Generation}} of {{Mobile Applications}}.
\newblock In {\em Proceedings of the 16th {{International Conference}} on
  {{Fundamental Approaches}} to {{Software Engineering}}}, FASE'13, pages
  250--265, Rome, Italy, 2013. {Springer-Verlag}.

\bibitem{Ye:MoMM'13}
H.~Ye, S.~Cheng, L.~Zhang, and F.~Jiang.
\newblock {{DroidFuzzer}}: {{Fuzzing}} the {{Android Apps}} with
  {{Intent}}-{{Filter Tag}}.
\newblock In {\em Proceedings of {{International Conference}} on {{Advances}}
  in {{Mobile Computing}} \&\#38; {{Multimedia}}}, MoMM '13, pages
  68:68--68:74, Vienna, Austria, 2013. {ACM}.

\bibitem{Zaeem:ICST'14}
R.~N. Zaeem, M.~R. Prasad, and S.~Khurshid.
\newblock Automated {{Generation}} of {{Oracles}} for {{Testing
  User}}-{{Interaction Features}} of {{Mobile Apps}}.
\newblock In {\em Proceedings of the 2014 {{IEEE International Conference}} on
  {{Software Testing}}, {{Verification}}, and {{Validation}}}, ICST '14, pages
  183--192, Washington, DC, USA, 2014. {IEEE Computer Society}.

\bibitem{Zhou:CIKM'12}
J.~Zhou and H.~Zhang.
\newblock Learning to {{Rank Duplicate Bug Reports}}.
\newblock In {\em Proceedings of the 21st {{ACM International Conference}} on
  {{Information}} and {{Knowledge Management}}}, CIKM '12, pages 852--861,
  Maui, Hawaii, USA, 2012. {ACM}.

\bibitem{Zhou:ICSE'12}
J.~Zhou, H.~Zhang, and D.~Lo.
\newblock Where {{Should}} the {{Bugs Be Fixed}}? - {{More Accurate Information
  Retrieval}}-based {{Bug Localization Based}} on {{Bug Reports}}.
\newblock In {\em Proceedings of the 34th {{International Conference}} on
  {{Software Engineering}}}, ICSE '12, pages 14--24, Zurich, Switzerland, 2012.
  {IEEE Press}.

\end{thebibliography}
\end{document}